\documentclass[conference]{IEEEtran}
\IEEEoverridecommandlockouts
\usepackage{cite}
\usepackage{amsmath,amssymb,amsfonts}
\usepackage{graphicx}
\usepackage{textcomp}
\usepackage{multicol}
\usepackage{multirow}
\usepackage{afterpage}
\usepackage{siunitx}
\usepackage{nccmath}
\usepackage{array}
\newcolumntype{M}[1]{>{\centering\arraybackslash}m{#1}}
\usepackage{makecell}
\usepackage{hhline}
\usepackage{booktabs}
\usepackage{xcolor}
\usepackage{xfrac,nicefrac}
\newcommand{\argmin}{\mathop{\mathrm{arg\,min}}}
\usepackage{newtxmath}
\def\BibTeX{{\rm B\kern-.05em{\sc i\kern-.025em b}\kern-.08em
    T\kern-.1667em\lower.7ex\hbox{E}\kern-.125emX}}
\usepackage[ruled]{algorithm2e}
\usepackage{algpseudocode}
\usepackage{stackengine}

\begin{document}
\title{Performance Comparison of Numerical Optimization Algorithms for {RSS}-{TOA}-Based Target Localization}

\makeatletter
\newcommand{\linebreakand}{%
  \end{@IEEEauthorhalign}
  \hfill\mbox{}\par
  \mbox{}\hfill\begin{@IEEEauthorhalign}
}
\makeatother

\author{\IEEEauthorblockN{Halim Lee}
\IEEEauthorblockA{\textit{School of Integrated Technology} \\
\textit{Yonsei University}\\
Incheon, Korea \\
halim.lee@yonsei.ac.kr} 
\and
\IEEEauthorblockN{Jiwon Seo} 
\IEEEauthorblockA{\textit{School of Integrated Technology} \\
\textit{Yonsei University}\\
Incheon, Korea \\
jiwon.seo@yonsei.ac.kr}
}

\maketitle

\begin{abstract}
In the literature, a hyper-enhanced local positioning system (HELPS) was developed to locate a target mobile device in an emergency.
HELPS finds the target mobile device (i.e., emergency caller) using multiple receivers (i.e., signal measurement equipment of first responders) that measure the received signal strength (RSS) and time of arrival (TOA) of the long-term evolution (LTE) uplink signal from the target mobile device.
The maximum likelihood (ML) estimator can be applied to localize a target mobile device using the RSS and TOA.
However, the ML estimator for the RSS-TOA-based target localization problem is nonconvex and nonlinear, having no analytical solution.
Therefore, the ML estimator should be solved numerically, unless it is relaxed into a convex or linear form.
This study investigates the target localization performance and computational complexity of numerical methods for solving an ML estimator.
The three widely used numerical methods are: grid search, gradient descent, and particle swarm optimization.
In the experimental evaluation, the grid search yielded the lowest target localization root-mean-squared error; however, the 95th percentile error of the grid search was larger than those of the other two algorithms.
The average code computation time of the grid search was extremely large compared with those of the other two algorithms, and gradient descent exhibited the lowest computation time.
HELPS can select numerical algorithms by considering their constraints (e.g., the computational resources of the localization server or target accuracy).
\end{abstract}

\begin{IEEEkeywords}
Target Localization, Numerical Algorithms, Emergency Response, E911 Positioning.
\end{IEEEkeywords}

\section{Introduction}

Target localization (i.e., passive target localization or passive emitter localization) \cite{Tomic19:A, Katwe20:NLOS, Wang13:Reference, Panwar22:A, Vaghefi12:Cooperative, Ouyang10:Received, Li07:Collaborative, Jeong20:RSS, Lee22:Evaluation, Lee23:Performance} is widely used in mission-critical applications, such as asset localization, wireless sensor networks, and search and rescue; target localization refers to the estimation of the location of a target device that transmits a wireless signal using a single or multiple receivers.

In an emergency response (e.g., ``911 emergency''), the target localization technologies are especially helpful in localizing the emergency caller.
Currently, several regions, such as the United States (U.S.A) and Europe, require mobile service providers to provide position information on the target mobile device in the emergency situations \cite{FCC15, EU14}.

In open-sky environments, the meter-level position information of the target mobile device can be provided using a global navigation satellite system (GNSS) \cite{Enge94:The, Dabove19:Towards, Yoon20:An}, such as global positioning service (GPS) of the U.S.A, GLONASS of Russia, and Galileo of Europe.
However, in harsh signal environments, such as dense urban, indoor, and tunnel environments, the accuracy and availability of GNSS-based positioning rapidly decrease because of signal blockage or reflection \cite{Lee22:Urban, Agarwal02:Algorithms, Lee22:Nonlinear, Lee19:Safety, Kim21:GPS, Kim14:Multi, Kang20:Practical, Lee22:Performance, Jia21:Ground, Lee20:Integrity}.
Further, GNSS signal is vulnerable to malicious jamming \cite{Park18:Dual, Park21:Single, Kim20:Development, Kim22:First, Park20:Effect, Rhee21:Enhanced, Lee22:SFOL} or ionospheric anomalies \cite{Lee22:Optimal, Sun21:Markov, Yoon14:Medium, Lee17:Monitoring, Bang13:Methodology}.

Target localization can be used as an alternative positioning method in such environments.
Recently, researchers in South Korea have developed a positioning system for emergency responses called the hyper-enhanced local positioning system (HELPS) \cite{Moon23:HELPS, Min22:Detection}, which applies the concept of target localization to its positioning algorithm.

Fig. \ref{fig:HELPS} shows the operational concept of the HELPS, which has four main components: a target mobile device, receivers (i.e., signal measurement equipment (MSE)), localization server (i.e., location calculation server (LCS)), and base station. 

\begin{figure}
    \centering
    \includegraphics[width=\linewidth]{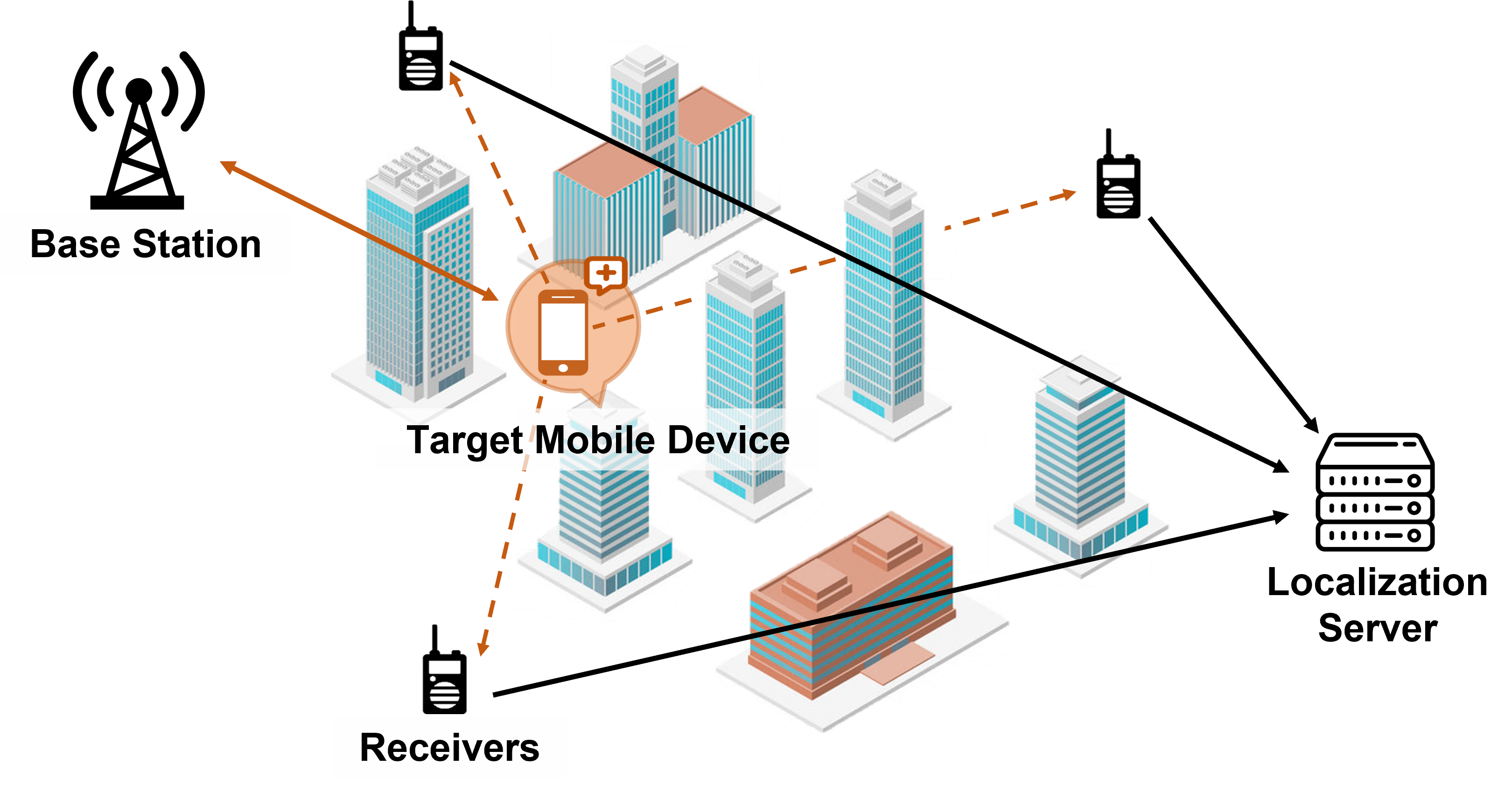}
    \caption{Operational concept of HELPS.}
    \label{fig:HELPS}
\end{figure}

When a target mobile device makes an emergency call, the emergency control center roughly identifies the position of the target mobile device using less accurate positioning methods (e.g., single-cellular-tower based localization) \cite{Zhao02:Standardization, Tsalolikhin11:A, Vossiek03:Wireless}.
Subsequently, the first responders are dispatched to the emergency site, and a long-term evolution (LTE) uplink signal, transmitted from the target mobile device using receivers, is measured.
These signal measurements are transmitted to a localization server, which estimates the position of the target mobile device.
This study focuses on the positioning algorithm of HELPS, which is performed out of the localization server.
Further details on HELPS can be found in \cite{Moon23:HELPS, Min22:Detection}.

Among several signal measurements, such as the received signal strength (RSS) \cite{Vaghefi12:Cooperative, Li07:Collaborative, Ouyang10:Received, Jeong20:RSS, Lee22:Evaluation, Lee23:Performance, Tomic19:A, Katwe20:NLOS, Wang13:Reference, Panwar22:A}, channel impulse response (CIR) \cite{Lee20:Preliminary, Lee20:Neural} and time-of-arrival (TOA) \cite{Tomic19:A, Katwe20:NLOS, Wang13:Reference, Panwar22:A}, this study assumes that receivers can measure the RSS and TOA because they can be obtained using relatively portable receivers.

Maximum likelihood estimation (MLE) \cite{Vaghefi12:Cooperative, Li07:Collaborative, Ouyang10:Received, Jeong20:RSS, Lee22:Evaluation, Lee23:Performance} can be applied to calculate the position of a target mobile device using both the RSS and TOA measurements.
However, the cost function of MLE for RSS-TOA-based target localization is nonconvex and nonlinear.
As the cost function of MLE has no closed-form solution, it cannot be solved analytically.

Previous studies \cite{Vaghefi12:Cooperative, Tomic19:A, Katwe20:NLOS, Wang13:Reference, Panwar22:A} proposed various optimization methods to obtain optimal or suboptimal solutions, and they considered wireless signal networks (WSNs) that have limited computational resources.
Therefore, they relaxed the original maximum likelihood (ML) estimator to a closed-form estimator, which can potentially lower the target localization performance when the RSS error is large.

This study simulates the emergency situations wherein the target must be precisely located, even if a relatively large amount of computational resources is required on the localization server.
Therefore, numerical algorithms were considered to find the optimal solution without relaxation of the ML estimator.

The three popular numerical algorithms are grid search \cite{Lee22:Evaluation, Lee23:Performance}, gradient descent \cite{Baldi95:Gradient, Ruder16:An}, and particle swarm optimization (PSO) \cite{Kulkarni10:Particle, Parvin19:Particle}.
These are well-used algorithms, but their performances in the TOA-RSS-based target localization problem have not been evaluated in the literature.
Therefore, this study experimentally evaluated the target localization performance and computation time of three algorithms for the RSS-TOA-based target localization problem.

The remainder of this paper is organized as follows.
Section \ref{sec:ProblemDescription} describes the RSS-TOA-based target localization problem and maximum likelihood (ML) estimator.
Section \ref{sec:OptimizationAlgorithms} briefly describes the three optimization methods used in this study: grid search, gradient descent, and PSO.
Section \ref{sec:ExperimentalSettings} introduces the experimental settings for evaluation and Section \ref{sec:ExperimentalResults} discusses the experimental results.
Finally, Section \ref{sec:Conclusion} concludes the paper.

\section{Problem Description} \label{sec:ProblemDescription}

\subsection{RSS and TOA Measurements} \label{subsec:RSSTOA}
When a total of $N$ receivers are dispatched to the emergency site, the RSS $P_i$ (dBm) measured by the $i$-th receiver can be modeled as a log-normal path loss model \cite{Vaghefi12:Cooperative, Ouyang10:Received, Li07:Collaborative, Lee22:Evaluation}.

\begin{equation} \label{eq:RSS}
\begin{split}
P_{i} &= P_{0} - 10 \beta \log_{10} \frac{d_i}{d_0} + n_{\mathrm{RSS}, i},\\
d_i &= \|\mathbf{x}-\mathbf{r}_i\|, \\
n_i &\sim \mathcal{N}(0, \sigma_\mathrm{dB}^2),
\end{split}
\end{equation}
where $P_{0}$ (dBm) is the signal power at the reference distance $d_0$ from the target ($d_0$ is set to 1 m in this study); $\beta$ is the path loss exponent; $n_{\mathrm{RSS}, i}$ is the log-normal shadowing term which is modeled as a zero-mean normal distribution with a standard deviation of $\sigma_\mathrm{RSS}$ (dB); $\mathbf{x}=[x_t,y_t]^\mathsf{T}$ is the 2D position of the target mobile device; $\mathbf{r}_i=[x_{r_i},y_{r_i}]^\mathsf{T}$ is the 2D position of $i$-th receiver; and $\|\cdot\|$ denotes $L^2$ norm.

The TOA $T_i$ (s) is measured by the $i$-th receiver, which can be modeled as

\begin{equation} \label{eq:TOA}
\begin{split}
T_{i} &= \frac{d_{i}}{c} + \tau + n_{\mathrm{TOA}, i},\\
\end{split}
\end{equation}
where $c$ is the speed of light; $\tau$ is the clock bias of the target mobile device; and $n_{\mathrm{TOA}, i}$ is the random noise term which is modeled as a normal distribution with a standard deviation of $\sigma_\mathrm{TOA}$ (s).

\subsection{Maximum Likelihood Estimator} \label{subsec:MLE}
In Eqs. (\ref{eq:RSS}) and (\ref{eq:TOA}), four unknowns should be estimated: $x_t$, $y_t$, $P_{0}$, and $\tau$.
The ML estimator for estimating the unknowns, $\boldsymbol{\thetaup} = [\mathbf{x}^\mathsf{T}, P_{0}, \tau]^\mathsf{T} = [x_t, y_t, P_{0}, \tau]^\mathsf{T}$, is given by

\begin{equation} \label{eq:MLE}
\begin{split}
    \widehat{\boldsymbol{\thetaup}} &= \argmin_{\boldsymbol{\thetaup}} \;\boldsymbol{F}\!\left(\boldsymbol{\thetaup}\right)\\
    &= \argmin_{\boldsymbol{\thetaup}} \, \sum_{i}^{N} \, \biggl\{ \left( P_{i} - P_{0} + 10 \beta \log_{10} d_i \right) ^2 \\ &\qquad\qquad\qquad\qquad + w \cdot \left( T_{i} - d_{i}/c - \tau \right) ^2 \biggr\},
\end{split}
\end{equation}
where $w$ is a weighting factor that considers the scale of RSS and TOA errors.
In this study, we empirically set the weighting factor as $w=4 \cdot 10^{-5} d_i - 10^{-3}$.

As previously mentioned, the ML estimator in Eq. (\ref{eq:MLE}) has no closed-form solution.
Therefore, the ML estimator in Eq. (\ref{eq:MLE}) should be solved numerically.

\section{Optimization Algorithms} \label{sec:OptimizationAlgorithms}

This section briefly introduces the three optimization algorithms used to obtain the numerical solution of Eq. (\ref{eq:MLE}).

\subsection{Grid Search} \label{subsec:GridSearch}
Grid search \cite{Lee22:Evaluation, Lee23:Performance} is a brute-force search, which finds the optimal solution that minimizes cost after sequentially evaluating every combination of parameters (i.e., candidate solutions).
The candidate solutions of the grid search are determined within the search space and at search interval.
The closer the search interval, the more accurate is the optical solution; however, more computation time is required.
A grid search is a simple and high-performance method that is computationally inefficient.

\subsection{Gradient Descent} \label{subsec:GradientDescent}
Gradient descent (i.e., steepest descent) \cite{Baldi95:Gradient, Ruder16:An} is an iterative algorithm that determines the local minimum by traveling forward in the opposite direction to the gradient.
Starting with the initial guess of the parameters $\boldsymbol{\thetaup}_0$, the next guess $\boldsymbol{\thetaup}_{k+1} \, (k=0,1,2,\cdots)$ is updated as follows \cite{Ruder16:An}:

\begin{equation} \label{eq:GradientDescent}
\begin{split}
    \boldsymbol{\thetaup}_{k+1} = \boldsymbol{\thetaup}_{k} - \gamma \cdot \nabla \boldsymbol{F}\!\left(\boldsymbol{\thetaup}_{k}\right),
\end{split}
\end{equation}
where $\gamma \!\in\! \mathbb{R}_{+}$ is the learning rate; and $\nabla \boldsymbol{F}\!\left(\cdot\right)$ denotes the gradient of function $\boldsymbol{F}$. If the learning rate $\gamma$ is set properly, $\boldsymbol{F}\!\left(\boldsymbol{\thetaup}\right)$ decreases gradually (i.e., $\boldsymbol{F}\!\left(\boldsymbol{\thetaup}_k\right)\geq\boldsymbol{F}\!\left(\boldsymbol{\thetaup}_ {k+1}\right)$).

\subsection{Particle Swarm Optimization} \label{subsec:PSO}
PSO \cite{Kulkarni10:Particle, Parvin19:Particle} is a population-based algorithm that determines the optimal solution by improving multiple candidate solutions (i.e., particles).
In every iteration, particles iteratively move to their local best position.
The final PSO solution has the lowest cost among the candidate solutions.
Because each agent optimizes its own position by exchanging information, it is expected to find the global minimum, even if some agents fall into the local minimum.

The PSO algorithm used in this study is implemented as shown in Algorithm \ref{alg:PSO}.
The maximum number of iterations is denoted as $M$ and total number of particles is denoted as $S$.
$\mathbf{b}_{\mathrm{lower}}$ and $\mathbf{b}_{\mathrm{upper}}$ are the lower and upper bounds on the parameters, respectively.
The final output of the algorithm is the best solution $\boldsymbol{\thetaup}_{\mathrm{best}}$.
Given the lower and upper bounds, the PSO algorithm initializes the population using a uniform distribution, and the velocity matrix is also randomly initialized.
At every epoch, the PSO algorithm updates the velocity of each particle, considering the best solution of the $i$-th particle (i.e., $\boldsymbol{p}_i$) and best solution of the entire population (i.e., $\boldsymbol{\thetaup}_\mathrm{best}$).
$w$, $c_1$, and $c_2$ are weights of each term that update the velocity.
After updating the position of the particle $\boldsymbol{\thetaup}_{i}$, $\boldsymbol{p}_i$ or $\boldsymbol{\thetaup}_\mathrm{best}$ can be updated if $\boldsymbol{\thetaup}_{i}$ has a lower cost compared with the current best solution.

\begin{algorithm}
\caption{PSO algorithm for estimating the best solution $\boldsymbol{\thetaup}_{\mathrm{best}}$}
\label{alg:PSO}
\vspace{0.2em}
\KwData{$\boldsymbol{F}$, $M$, $S$, $\mathbf{b}_{\mathrm{lower}}$, $\mathbf{b}_{\mathrm{upper}}$}
\KwResult{$\boldsymbol{\thetaup}_{\mathrm{best}}$}
Randomly initialize the population $\mathcal{P} = [ \boldsymbol{\thetaup}_1, \boldsymbol{\thetaup}_2, \cdots, \boldsymbol{\thetaup}_{S}]$ within $\left[\mathbf{b}_{\mathrm{lower}}, \mathbf{b}_{\mathrm{upper}}\right]$ \\
Randomly initialize the velocity $\mathcal{V} = [\boldsymbol{v}_1,\boldsymbol{v}_2, \cdots, \boldsymbol{v}_s]$

 \For{\upshape $iter \in {1,2,\cdots,M}$}{
 \For{\upshape $i \in {1,2,\cdots,S}$}{
 Pick random numbers $r_1$, $r_2$ within $[0,1]$
 $\boldsymbol{v}_i \gets w \cdot \boldsymbol{v}_i + c_1 \cdot r_1 \cdot (\boldsymbol{p}_i-\boldsymbol{\thetaup}_i) + c_2 \cdot r_2 \cdot (\boldsymbol{\thetaup}_{\mathrm{best}}-\boldsymbol{\thetaup}_i)$
 $\boldsymbol{\thetaup}_i \gets \boldsymbol{\thetaup}_i + \boldsymbol{v}_i$

 \If{\upshape $\boldsymbol{F}\!(\boldsymbol{\thetaup}_i) < \boldsymbol{F}\!(\boldsymbol{p}_i)$} {
 $\boldsymbol{p}_i \gets \boldsymbol{\thetaup}_i$ \\
 \If{\upshape $\boldsymbol{F}\!(\boldsymbol{p}_i) < \boldsymbol{F}\!(\boldsymbol{\thetaup}_\mathrm{best})$} {
 $\boldsymbol{\thetaup}_\mathrm{best} \gets \boldsymbol{p}_i$
 }
 }
 }
 }
\Return $\boldsymbol{\thetaup}_\mathrm{best}$
\vspace{0.2em}
\end{algorithm}

\section{Experimental Settings} \label{sec:ExperimentalSettings}

Fig. \ref{fig:ExperimentalEnvironment} shows the experimental environment near the Seongdong Bridge, Seoul, South Korea.
The target mobile device was placed below the bridge, as indicated by the red phone icon.
Four signal collection points, 50, 100, 150, and 200 m from the target mobile device, are indicated by the yellow person icons.
The RSS and TOA were measured 90 times at each signal-collection point.
A Samsung Galaxy S8+ was used as the mobile device, Xilinx XCKU9P 1FFVE900I was used as the receiver, and field-programmable gate array (FPGA) board was programmed to measure the RSS and TOA of the signal transmitted from the target mobile device \cite{Moon23:HELPS, Min22:Detection}.
The central frequency of the signal was set to 738 MHz, and bandwidth was set to 10 MHz.

In the grid search, the search space for $\mathbf{x}$ was set to twice the distance between the target mobile device and  receiver (i.e., the search space was set to 100, 200, 300, and 400 m for each experimental setting in Fig. \ref{fig:ExperimentalEnvironment}).
The search interval was set at 1 m. 
The search space and search interval for $P0$ were set to 6 and 0.5 dB, respectively.
The search space and search interval for $c\cdot\tau$ were set to 50 and 5 m, respectively.
The initial values of parameters were obtained by semidefinite programming (SDP) in \cite{Vaghefi12:Cooperative}.
The learning rate of the gradient descent was set to 0.001.
The initial guess of the parameters was set as $\boldsymbol{\thetaup}_0 = [x_t-20, y_t-20, -60, 1350]^\mathsf{T}$.
The total iteration of the gradient descent was set to 200.
In the PSO, $M$ was set to 200 and $S$ was set to 100.
The weights $w$, $c_1$, and $c_2$ were set to 0.8, 0.1, and 0.1, respectively.
The lower and upper bounds ($b_\mathrm{lower}$ and $b_\mathrm{upper}$) were set as in the search space of the grid search.

Algorithms were programmed in Python and ran on a desktop with an Intel Core i7-8700 CPU operating at 3.20 GHz and 16 GB of memory.

\begin{figure}
    \centering
    \includegraphics[width=\linewidth]{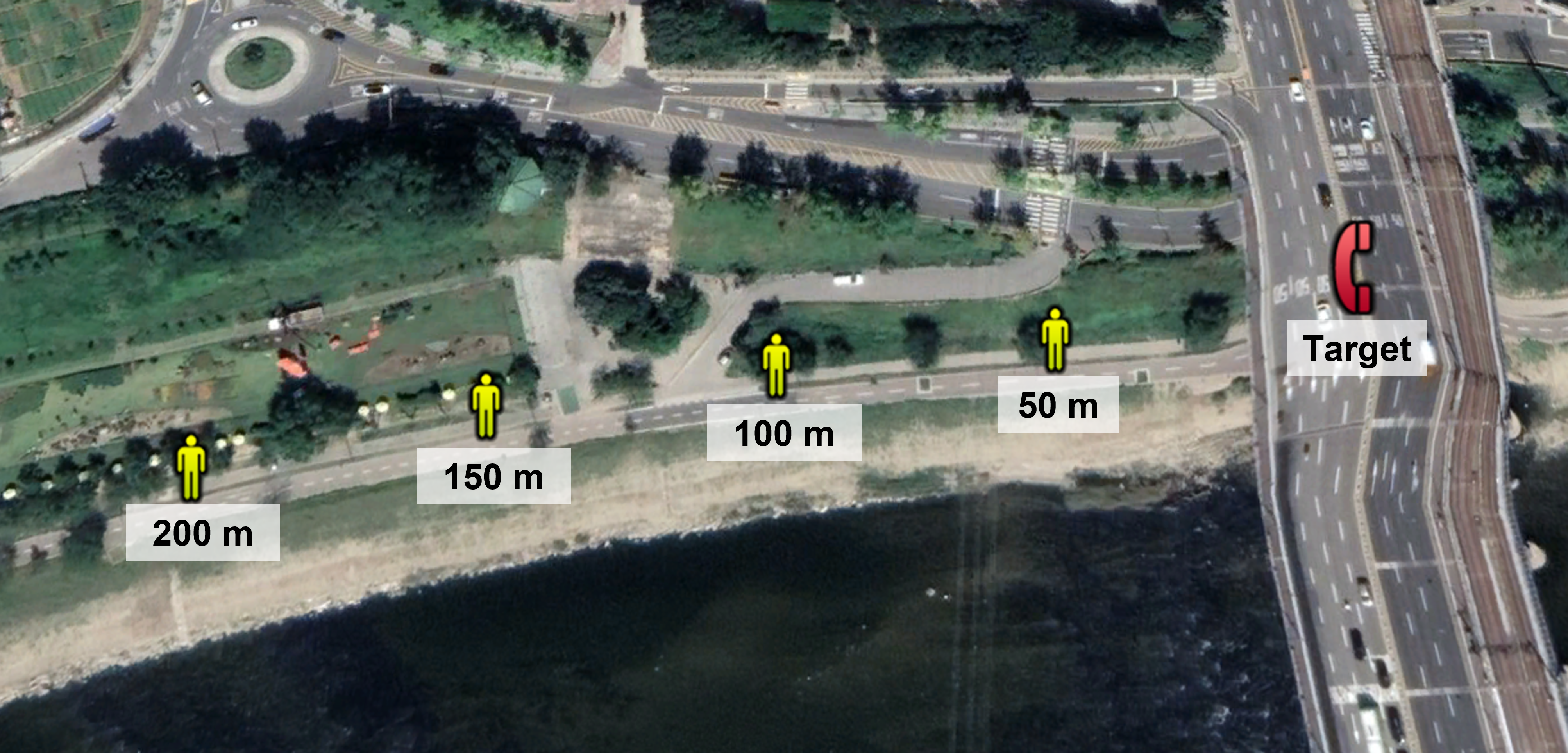}
    \caption{Experimental environment.}
    \label{fig:ExperimentalEnvironment}
\end{figure}

\section{Experimental Results} \label{sec:ExperimentalResults}

\subsection{Performance Comparison of Three Algorithms} \label{subsec:PerformanceComparison}
Fig. \ref{fig:ExperimentalResult} shows a 2D plot and cumulative distribution function (CDF) of the target localization errors of the three algorithms.
For the evaluation, it was assumed that the four receivers were placed evenly around the target mobile device.
The root mean square errors (RMSE) of the grid search, gradient descent, and PSO algorithms are 18.50, 22.29, and 20.37 m, respectively.
When comparing the RMSE, the grid search showed better performance than the other two algorithms.

Table \ref{tab:Performance} compares the 80th and 95th percentiles of the target localization errors for the three algorithms.
In emergencies, it is important to estimate the locations of emergency callers without omissions. 
Therefore, ensuring a specific level of target localization accuracy for strict criteria, such as 80\% or 95\%, is necessary. 
When comparing the 95th percentile error, grid search showed poor performance compared with gradient descent or PSO.
However, we assumed that the initial estimate was fairly close to the global minimum (i.e., the initial estimate was approximately 28.3 m away from the global minimum) in the gradient descent algorithm.
Therefore, the performance of gradient descent is expected to worsen if the initial guess is far from the global minimum.

\begin{figure}
    \centering
    \includegraphics[width=0.85\linewidth]
    {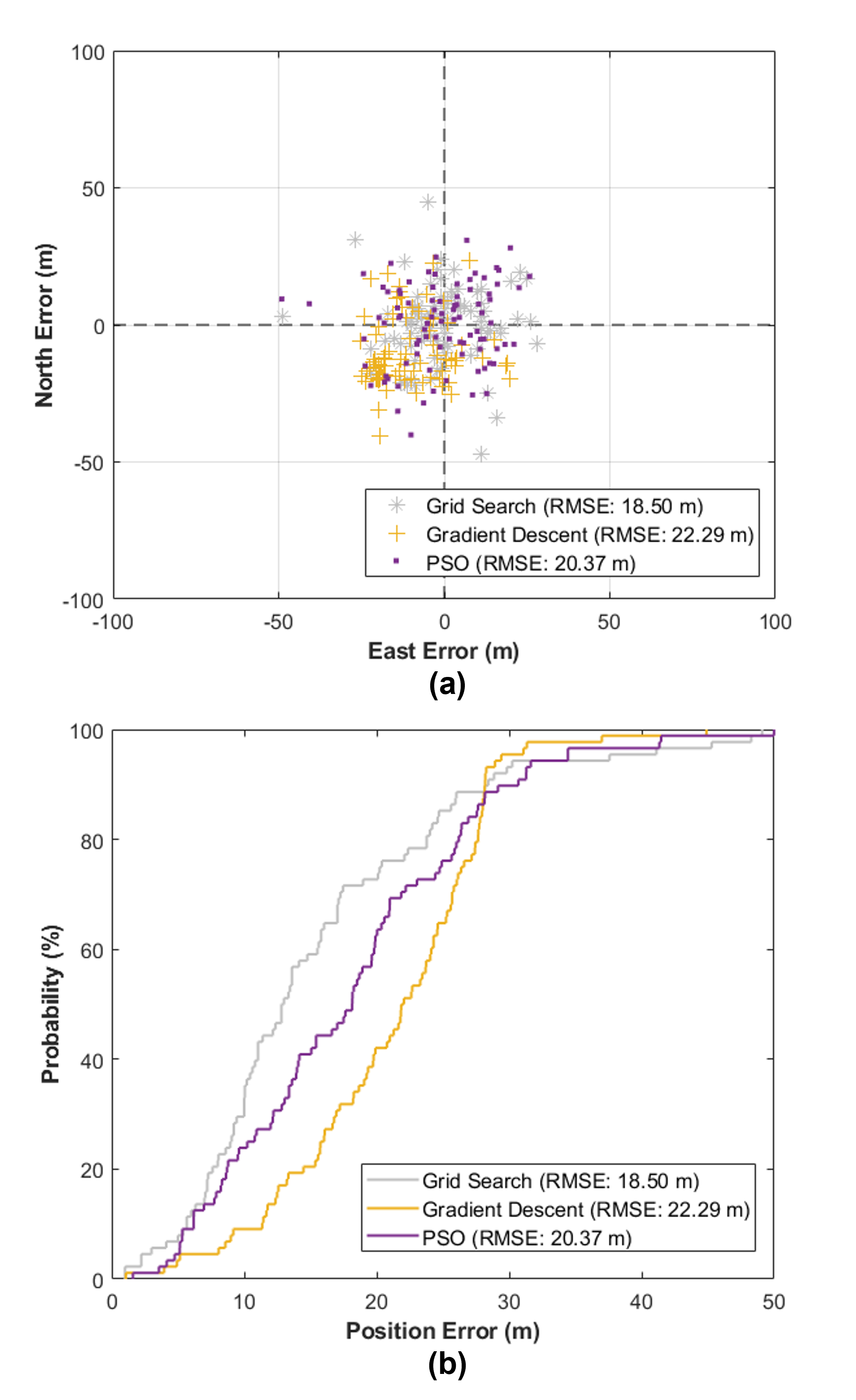}
    \caption{(a) 2D target localization error and (b) cumulative localization error distribution of the three algorithms.}
    \label{fig:ExperimentalResult}
\end{figure}

\begin{table}
\small
\centering
\caption{Performance comparison among three algorithms (unit: m)}
\label{tab:Performance}
\vspace{-4mm}
\begin{center}
{\renewcommand{\arraystretch}{1.4}
 \begin{tabular}[c]{M{0.9cm} M{1.9cm} M{2.1cm} M{1.7cm}}
 \toprule
    {} & \textbf{Grid Search} 
       & \textbf{Gradient Descent} 
       & \textbf{PSO} \\
    \noalign{\vspace{2pt}}
\hline
\noalign{\vspace{2pt}}
 80\% & 23.77 & 27.65 & 26.20 \\
 95\% & 37.58 & 29.40 & 34.43 \\
 \noalign{\vspace{1pt}}
 \bottomrule
\end{tabular}}
\end{center}
\end{table}

\subsection{Comparison of Computation Time} \label{subsec:ComparisonComplexity}
Fig. \ref{fig:CodeTime} compares the average code computation times of the three algorithms.
The grid search is computationally inefficient compared with  the other two algorithms.
As the search space increased (i.e., the distance between the target and receiver increased), the code computation time increased significantly.
The computation time of the grid search can be reduced by increasing the search interval and decreasing the search space; however, there is a risk of increasing the target localization error.
In the case of PSO, the computation time can be reduced by reducing the number of particles and total epochs.

\begin{figure}
    \centering
    \includegraphics[width=0.85\linewidth]{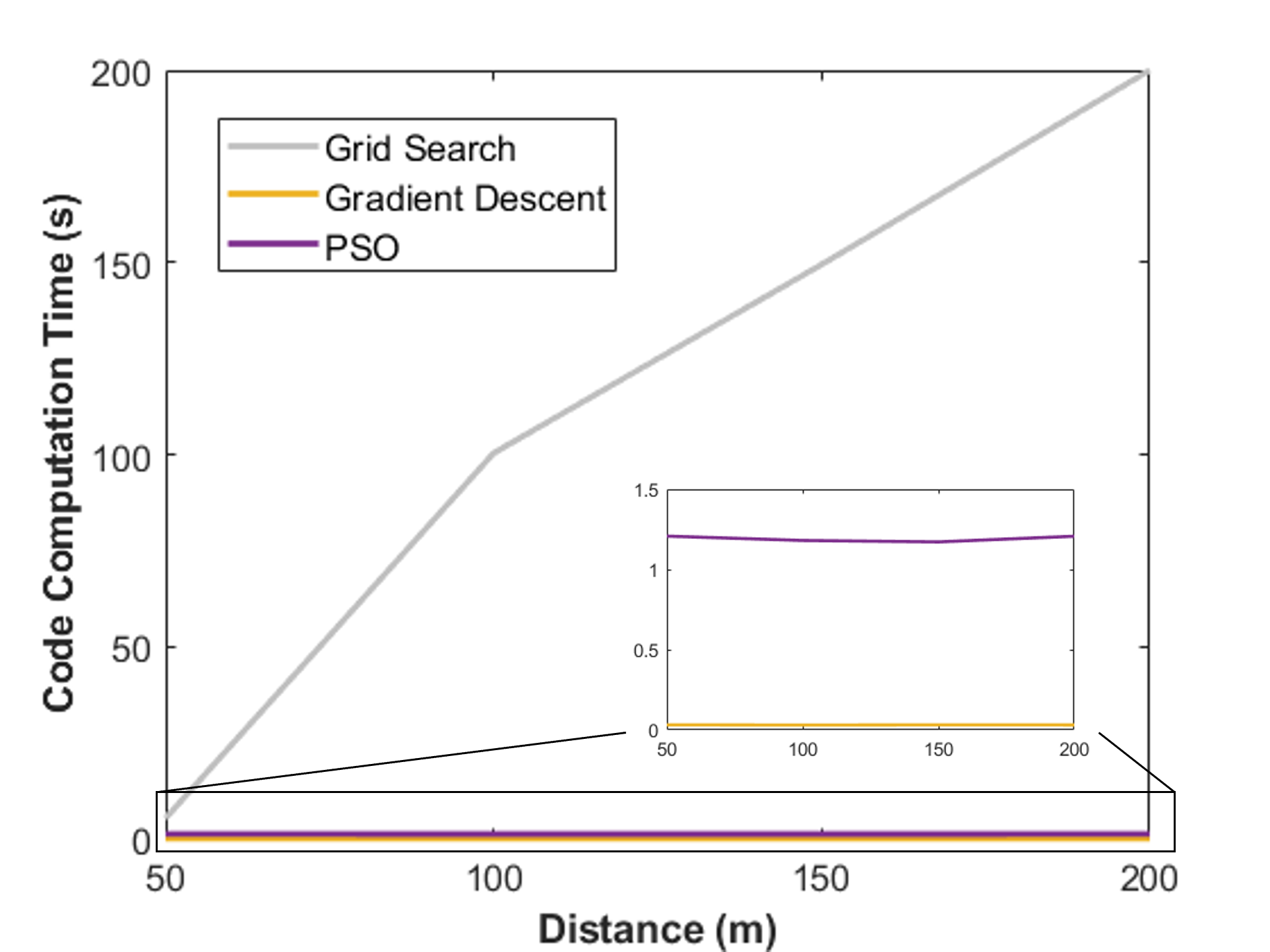}
    \caption{Average code computation time of the three algorithms.}
    \label{fig:CodeTime}
\end{figure}

\section{Conclusion} \label{sec:Conclusion}
This study solved the nonconvex and nonlinear ML estimators for RSS-TOA-based target localization and considered three numerical methods for solving the ML estimator: grid search, gradient descent, and PSO.
We experimentally evaluated the target localization performances and computational complexities of the three algorithms.
The target localization RMSE decreases in the following order: grid search, PSO, and gradient descent.
However, the 95th percentile error of the grid search was larger than those of the other two methods.
The average code computation time increases in the following order: grid search, PSO, and gradient descent.
The grid search was computationally inefficient compared with the other two algorithms.

\section*{Acknowledgment}

The authors would like to thank the Communication System Laboratory of Hanyang University, Korea, for the data collection.
This research was supported in part by the Institute of Information \& Communications Technology Planning \& Evaluation (IITP) grant funded by the Korea government (KNPA) (2019-0-01291); in part by the Future Space Navigation \& Satellite Research Center through the National Research Foundation funded by the Ministry of Science and ICT, Republic of Korea (2022M1A3C2074404); and in part by the Unmanned Vehicles Core Technology Research and Development Program through the National Research Foundation of Korea (NRF) and the Unmanned Vehicle Advanced Research Center (UVARC) funded by the Ministry of Science and ICT, Republic of Korea (2020M3C1C1A01086407).

\bibliographystyle{IEEEtran}
\bibliography{mybibfile, IUS_publications}

\end{document}